\documentclass[aps,prl,twocolumn,groupedaddress]{revtex4-1}
\usepackage{graphicx}
\usepackage{epsfig}
\usepackage{epstopdf}

\begin{document}
\title{Giant Cross Kerr Effect for Propagating Microwaves Induced by an Artificial Atom}

\author{Io-Chun Hoi$^{1}$, C.M. Wilson$^{1*}$, G\"oran Johansson$^{1}$, Tauno Palomaki$^{1}$, Thomas M. Stace$^{2}$, Bixuan Fan$^{2}$ \& Per Delsing$^{1}$}

\email{per.delsing@chalmers.se or chris.wilson@chalmers.se}

\affiliation{$^1$Department of Microtechnology and Nanoscience (MC2), Chalmers University of Technology, SE-412 96 G\"oteborg, Sweden}
\affiliation{$^2$Centre for Engineered Quantum Systems, School of Physical Sciences, University of Queensland, Saint Lucia, Queensland 4072, Australia}

\date{\today}

\begin{abstract}

We have investigated the cross Kerr phase shift of propagating microwave fields strongly coupled to an artificial atom.  The artificial atom is a superconducting transmon qubit in an open transmission line. We demonstrate average phase shifts of 11 degrees per photon between two coherent microwave fields both at the single-photon level.  At high control power, we observe phase shifts up to 30 degrees.  Our results provide an important step towards quantum gates with propagating photons in the microwave regime. 

\end{abstract}
 \pacs{}          
 \maketitle     
In recent years, there has been great interest in using photons as quantum bits for quantum information processing \cite{Knill}. The implementation of quantum logic gates using photons requires interactions between two fields \cite{Knill,Shapiro1}. One possible coupling mechanism is the Kerr effect, where the photons interact via a nonlinear medium. By means of the Kerr effect, quantum logic operations such as the controlled phase gate \cite{Turchette}, the quantum Fredkin gate \cite{Milburn} and the conditional phase switch \cite{Resch} can be realized. Moreover, for a sufficiently strong nonlinearity, quantum nondemolition detection of propagating photons may be possible by measuring the Kerr phase shift. Superconducting qubits provide a very strong nonlinearity \cite{Astafiev1,Hoi} that might be suitable for this purpose.  

In cavity QED experiments, Kerr phase shifts on the order of 10 degrees have been measured at the single-photon level \cite{llya}. However, in such configuration, the presence of the cavity limits the bandwidth, which constrains its usefulness over a wide range of frequencies. Therefore an open quantum systems without a cavity is advantageous. An example of such a system is atoms coupled to a 1D electromagnetic environment. A Kerr phase shift is also present in these systems, but so far the measured phase shift has been very small. In nonlinear photonic crystal fibers, for instance, an average Kerr phase shift of $10^{-6}$ degrees per photon has been measured \cite{Nobuyuki}. 

A new class of open quantum systems have been made possible by progress in circuit QED, providing a fascinating platform for engineering light-atom interactions \cite{Nakamura,Vion,Wallraff,Hofheinz,Chris1,Sandberg,Matthias} and testing fundamental aspects of quantum physics \cite{Chris2}.  In this letter, we embed a single artificial atom in an open transmission line \cite{Astafiev1,Hoi,Hoi2}. Through strong coupling, we achieve average phase shifts up to 11 degrees per photon between two coherent microwave fields at the single-photon level. This is six orders of magnitude larger than in optical systems \cite{Nobuyuki}. The Kerr effect demonstrated here also differs greatly from that previously demonstrated in superconducting devices. The origin of our Kerr effect is via a three-level artificial atom as opposed to the kinetic inductance of a superconducting film \cite{Erik} or the Josephson inductance of a Superconducting Quantum Interfere Device (SQUID) \cite{Castellanos}. Both of these Kerr media require a pump tone at least several orders of magnitude higher than the fields required using our three-level artificial atom.

\begin{figure}
\includegraphics[width=\columnwidth]{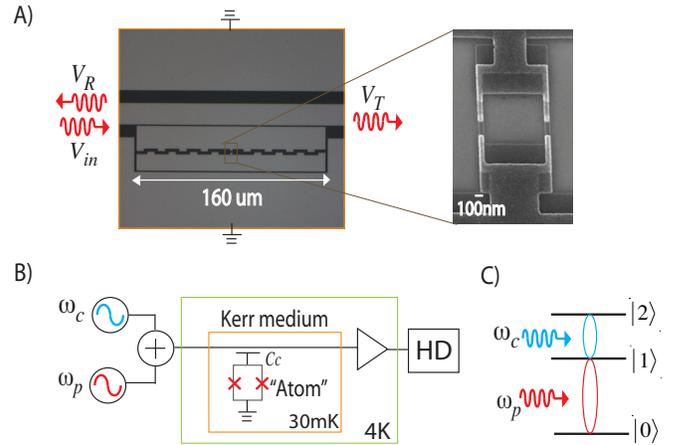}
\caption{A  micrograph of our artificial atom, a superconducting transmon qubit, embedded in a 1D open transmission line. The artificial atom acts as a Kerr medium. (Zoom In) Scanning-electron micrograph of the SQUID loop of the transmon, which allows us to tune the transition frequency of the transmon with an external magnetic flux, $\Phi$. (B) Schematic of the measurement setup using heterodyne detection (HD). C) The three-level artificial atom driven by a probe (red) and a control (blue) tone.}
\end{figure}

Our artificial atom is a superconducting transmon \cite{Koch}, strongly coupled to a $Z_0= 50$ $\Omega$ 1D open transmission line through a capacitance, $C_c$ (see Fig.\ 1A, B). The transition frequency between the ground state, $\left\vert 0 \right\rangle$, and the first excited state, $\left\vert 1 \right\rangle$, is $\omega_{01}(\Phi)/2\pi\sim 7.1$ GHz. An external magnetic flux $\Phi$ allows us to tune the transition frequency. The transition frequency between the first excited state and second excited state is $\omega_{12}/2\pi\sim 6.4$ GHz.

\begin{figure*}
\includegraphics[width = 2\columnwidth]{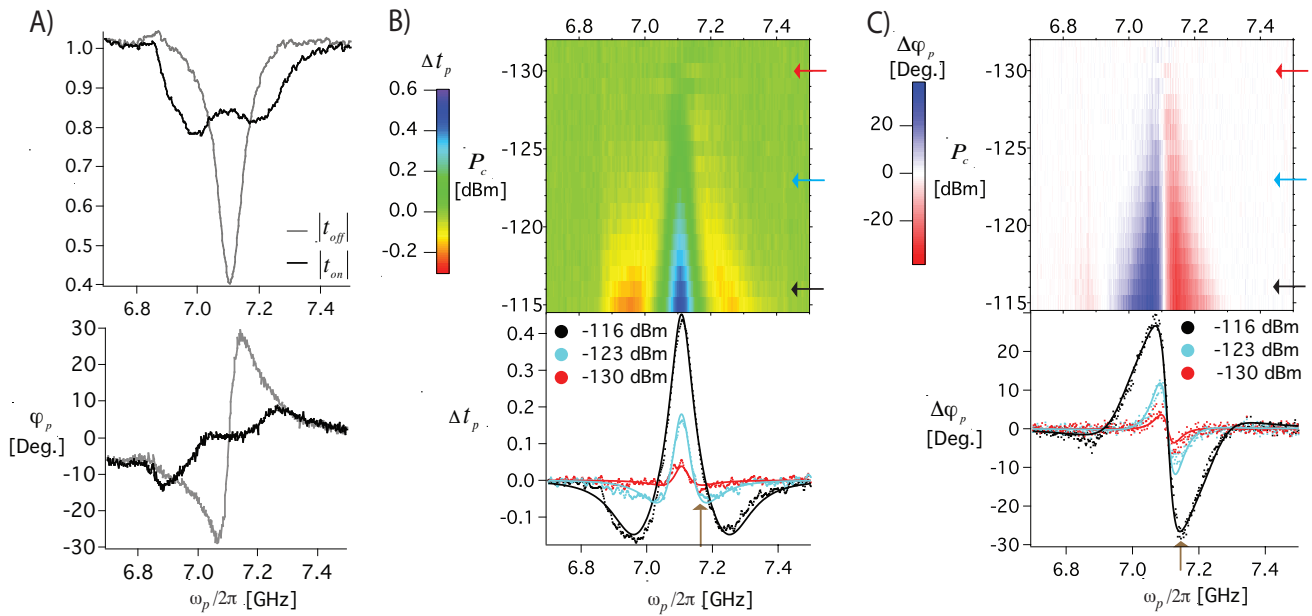}
\caption{Transmission coefficient for the probe, as a function of $\omega_{p}$ and control power, $P_c$, for low probe powers, $\Omega_p<<\gamma_{10}$. A) Top: Transmission coefficient with control power on, $|t_{on}|$ ($P_c = -116$ dBm), and off, $|t_{off}|$. Bottom: the corresponding phase response. B) Induced amplitude response, $\Delta t_{p}$. Top: Experimental data, Bottom: 3 horizontal line cuts. Curves represent the corresponding theory fits. C) Corresponding phase response, $\Delta\varphi_p$. Top: Experimental data, Bottom: 3 horizontal line cuts. The solid curves represent the corresponding theory fits. Brown arrows show the frequency that gives the maximum phase response and the corresponding amplitude response. We see that the maximum $\Delta\varphi_p$ corresponds to a small $\Delta t_p$. Note that we fit all of the data in amplitude and phase simultaneously.}
\end{figure*}

The electromagnetic field in the transmission line can be described by an incoming voltage wave, $V_{in}$, a transmitted wave, $V_{T}$, and a reflected wave, $V_{R}$. In Fig.\ 1A, the transmission and reflection coefficients are defined as $t=V_T/V_{in}$ and $r=V_R/V_{in}$, respectively. A coherent probe signal will drive coherent oscillations of the 0-1 atomic dipole at a Rabi frequency, $\Omega_p$, which is linear in the probe amplitude. For a weak resonant probe, where $\Omega_p$ is much less than the decoherence rate $\gamma_{10}$ of the 0-1 transition $(\Omega_p<<\gamma_{10})$, the forward propagating wave is completely reflected \cite{Shen2,Astafiev1,Hoi}. This effect can be understood as an interference effect between the incoming and
scattered waves. In general, we do not observe perfect extinction of the transmitted
field, with the residual transmission determined by the ratio of the pure dephasing
rate, $\Gamma_{\phi}$, to the relaxation rate $\Gamma_{10}$ \cite{Astafiev1,Hoi}. In related work, we have
shown that, due to the strong nonlinearity of the atomic response, the reflected field
is nonclassical, exhibiting photon antibunching \cite{Hoi2, Chang}. Assuming the only relaxation channel is emission into the transmission line, we have $\Gamma_{10}=\omega_{01}^2C_c^2Z_0/(4C_\Sigma)$, where $C_\Sigma$ is the total capacitance of the transmon. 

 \begin{figure}
 \includegraphics[width =3.2in]{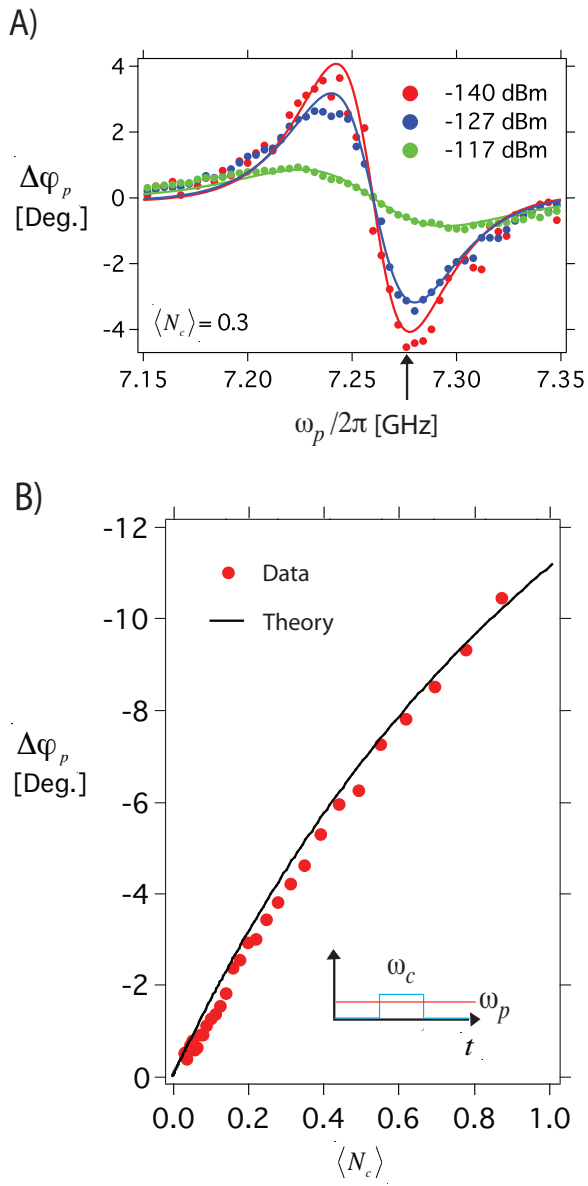}
\caption{Probe phase shift induced by a weak control pulse. A) $\Delta\varphi_p$ as a function of $\omega_{p}$ for three different probe powers and $\left \langle N_c \right \rangle\simeq0.3$. B) $\Delta\varphi_p$ as a function of $\left \langle N_c \right \rangle$ for a weak probe at $\delta\omega_p\simeq20$ MHz (indicated by the black arrow in Fig. 3B, where the phase shift is maximum). An average phase shift of 11 degrees per control photon is observed. Inset: The control pulse is used to change the phase of the continuous probe in the time domain. The solid curves in both panels are theoretical fits to the data using the parameters: $\Gamma_{10}/2\pi=140$ MHz, $\gamma_{10}/2\pi=100$ MHz, $\Gamma_{21}/2\pi=170$ MHz and $\gamma_{21}/2\pi=184$ MHz.}
\end{figure} 

By measuring the transmission coefficient as a function of probe frequency, $\omega_p$, at low probe power, $P_p$, the extinction dip provides the 0-1 transition frequency, $\omega_{01}/2\pi\sim 7.1$ GHz. The 1-2 transition can be directly measured using 2-tone spectroscopy \cite{Hoi}. We extract $\omega_{12}/2\pi=6.38$ GHz, giving an anharmonicity of $\alpha = 720$ MHz between the two transitions. As illustrated in Fig.\ 1B, C, we apply two continuous tones, the probe at $\omega_{p}\sim\omega_{01}$ and the control at $\omega_{c}=\omega_{12}$, and observe the induced amplitude and phase shift of the probe. These responses depend on four different parameters: the powers and the detunings of the probe and the control tones. In the following, we study the dependence of these responses on the two different powers and the probe detuning. We note that the dipole moment of the 1-2 transition is $\sqrt{2}$ larger than that of the 0-1 transition \cite{Koch}. Therefore, for a particular microwave power $P$, we have $\Omega_c(P)=\sqrt{2}\Omega_p(P)$.

In Fig.\ 2A, we define the induced transmission, $\Delta t_{p}$, as the difference between the magnitude of the probe transmission with control on, $t_{on}$, and control off, $t_{off}$, $\Delta t_{p}=|t_{on}|-|t_{off}|$. In Fig.\ 2A, we show $ t_{on}$, demonstrating the Aulter-Townes doublet \cite{Autler,Abdumalikov}. The doublet states appear as minima in the black curves of Fig.\ 2A with a separation given by $\Omega_c$. In Fig.\ 2B, we show $\Delta t_{p}$ as a function of $P_c$ and $\omega_{p}$ for low probe powers, $\Omega_p<<\gamma_{10}$. While the maximum induced amplitude response occurs when the probe is on resonance, the induced phase response is greatest at $\delta\omega_p\simeq20$ MHz, indicated by the brown arrow in Fig.\ 2C. Here we find phase shifts $\Delta\varphi_p$ up to 30 degrees for large control powers. Crucially, at this detuning the giant phase shift occurs dispersively, i.e., $\Delta t_{p}\simeq 0$.   

We model the system as a three-level atom with two drives (probe and control). The time evolution of the density matrix, $\rho$, is then given by the von Neumann equation. Taking into account the decoherence rates, we numerically compute the steady-state solution \cite{Abdumalikov}. The transmission coefficient for the probe can be written in terms of the off-diagonal elements, $\rho_{10}$ and $\rho_{01}$, using the quantum circuit represented in Fig.\ 1B. By fitting the measured amplitude and phase response as a function of probe frequency and control power simultaneously, and using the previous values, $\omega_{01}/2\pi=7.1$ GHz, $\omega_{12}/2\pi=6.38$ GHz, we extract the following parameters: $\Gamma_{10}/2\pi=74$ MHz and $\gamma_{10}/2\pi=60$ MHz, $\gamma_{21}/2\pi=90$ MHz. 
                                       
Using our Kerr medium in quantum application will require large phase shifts even with the control and probe at the single-photon level. Therefore, we are interested in the cross Kerr phase shift in the limit of low control power \cite{Schmidt}. Typically, the Kerr phase shift is expressed as $\Delta\varphi_p=kP_c$ at low control power, where $k$ is the Kerr coefficient. $P_c$ can be expressed in terms of the average number of control photons $\left \langle N_c \right \rangle$ per interaction time, $P_c=\left \langle N_c \right \rangle\hbar\omega_c(\Gamma_{21}/2\pi)$, with $P_c\simeq-121.4$ dBm $\simeq$ 0.72 fW corresponding to $\left \langle N_c \right \rangle=1$. Therefore, $\Delta\varphi_p$ is proportional to $\left \langle N_c \right \rangle$ at low $P_c$. As indicated by the red curve of Fig.\ 2C, $\Delta\varphi_p$ is several degrees when $P_c\simeq-130$ dBm. To measure this phase shift more accurately, we send a weak control pulse (see inset of Fig.\ 3B) and measure the phase difference of the probe when the control is on and off. Note that here the 0-1 transition is 7.26 GHz (due to a different external magnetic flux, $\Phi$). In Fig.\ 3A, we measure $\Delta\varphi_p$ as a function of probe frequency at $\left \langle N_c \right \rangle\simeq0.3$, ($P_c\simeq-127$ dBm). The maximum phase shift occurs at 7.28 GHz, detuned from the 0-1 resonance frequency by $\delta\omega_p\simeq20$ MHz. We then measure $\Delta\varphi_p$ at 7.28 GHz as a function of $\left \langle N_c \right \rangle$, as shown in Fig.\ 3B. We observe a linear dependence in the low control photon limit, with an slope of 11 degrees per control photon. This large phase shift per photon may enable quantum nondemolition measurement of propagating photons if incorporated with a quantum-limited amplifier \cite{clerk}. The noise temperature of our detection chain is around 7 K (30 photons). We averaged 2 million pulses for every data point in Fig.\ 3B.

\begin{figure}
 \includegraphics[width =3in]{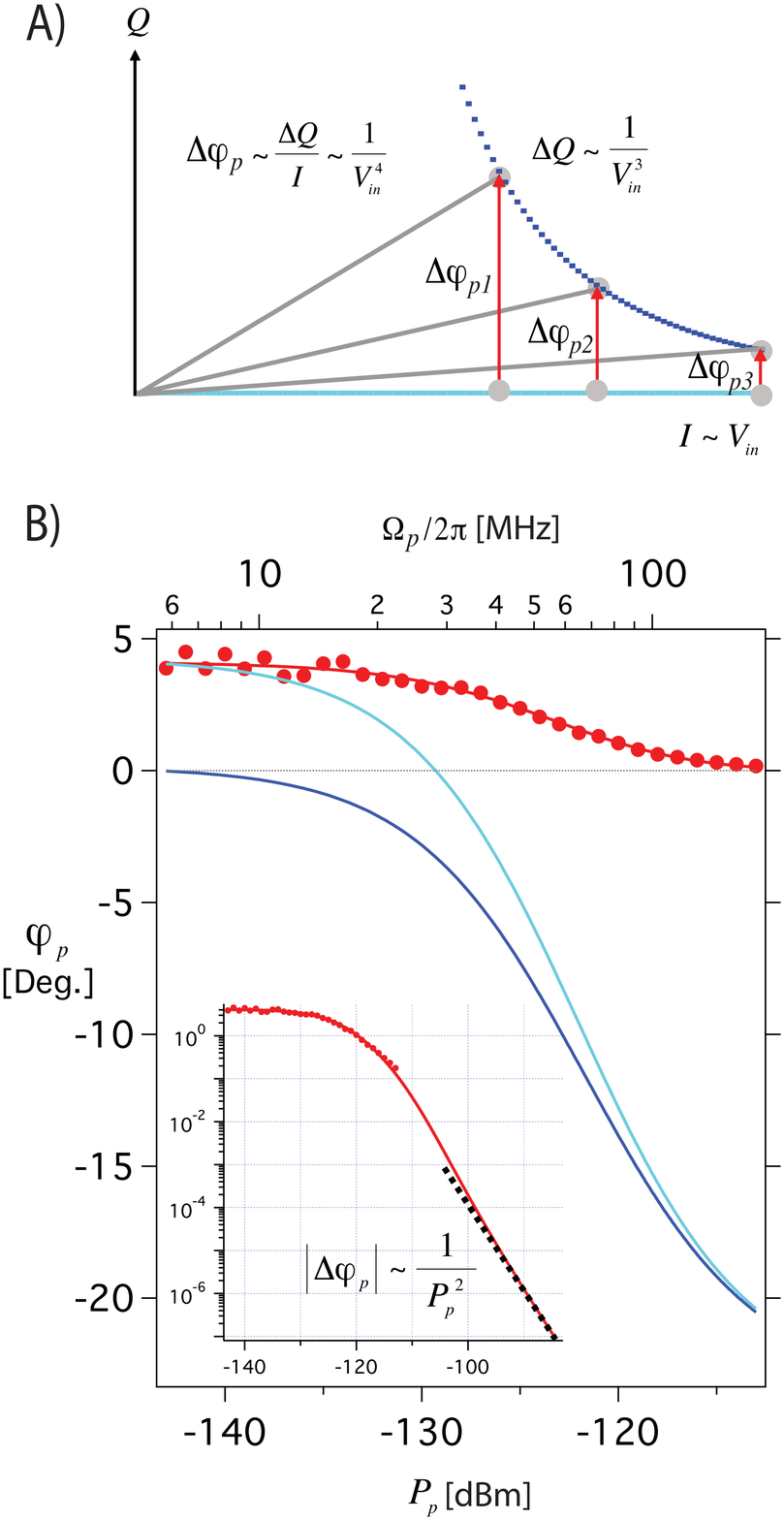}
\caption{Probe power dependence of $\Delta\varphi_p$ at $\left \langle N_c \right \rangle\simeq0.3$. A) Cartoon of the saturation of $\Delta\varphi_p$ in the phase space picture. I and Q denote the in-phase and quadrature component respectively. See text for the detailed explanations. B) The deep and light blue curves indicate the theoretical response of the probe phase with and without control field, respectively. $\Delta\varphi_p$ is the difference between the deep blue and light blue curves. Note that we introduce a -23.4 degrees offset in the two blue curves for display purposes. The red data shows $\Delta\varphi_p$ as a function of $P_p$ at $\delta\omega_p\simeq20$ MHz (indicated by black arrow in Fig.\ 3A). We see that $\Delta\varphi_p$ gradually becomes weaker as $P_p$ increases, saturating the 0-1 transition. $P_p\simeq-122$ dBm corresponds to $\left \langle N_p \right \rangle=1$. Inset:  Extension of the plot to higher powers.}
\end{figure} 

Fig.\ 4B shows the dependence of $\Delta\varphi_p$ on $P_p$ at $\left \langle N_c \right \rangle\simeq0.3$. The solid red curve is a theoretical calculation, with the same parameters as in Fig.\ 3. As expected, for high probe power, the probe phase response becomes weaker, and finally disappears due to the saturation of the artificial atom. This saturation behavior can be explained by the phase space cartoon in Fig.\ 4A. The in-phase component, $I$, is proportional to the amplitude of the probe, $V_{in}$. The quadrature component, $Q$, is proportional to $\rho_{10}$ (which is equal to $\rho_{01}$). In the case without the control field, $\rho_{10}$ is inversely proportional to the amplitude of the probe \cite{Astafiev1}. To reveal the saturation effect, we look at $\Delta\varphi_p$ (red curve), i.e. the difference between the deep blue and light blue curves. The difference between these two off-diagonal elements of the density matrix (same as $\Delta Q$), with and without a control field, is found to be inversely proportional to the cube of $V_{in}$, as indicated by the purple dotted curve of Fig.\ 4A. The $V_{in}^{-3}$ dependence (as $V_{in}\to \infty$) comes about from two different effects. Firstly, the probe transition becomes dressed, giving rise to an Autler-Townes splitting, which goes as $V_{in}$. This leads to a proportional detuning of the control field, and as a result of this detuning, the control field induces a population transfer, $\delta\rho$, between the states of the control transition that varies as $V_{in}^{-2}$. Further, via the Bloch equations, the displacement of the probe with and without the control field is proportional $\delta\rho/V_{in}$. In combination, we see that the probe displacement varies as $V_{in}^{-3}$, as seen in both the experiment and the full theoretical calculation. Therefore, $\Delta\varphi_p\sim V_{in}^{-4}\sim P_p^{-2}$, consistent with the slope of the red curve in the inset of Fig.\ 4B. 

In conclusion, we have observed an average cross Kerr phase shift of 11 degrees per photon between two coherent microwave fields at the single-photon level.  This system has the advantage of allowing photons to propagate over a wide range of frequencies compared to cavity-based systems \cite{Stojan}. Such giant Kerr phase shifts may find applications in quantum information processing.


We acknowledge financial support from the Swedish Research Council, the Wallenberg foundation, STINT and from the EU through the ERC and the projects SOLID and PROMISCE. TMS was funded by ARC EQuS. We would also like to acknowledge G. J. Milburn for fruitful discussions.

%

\end{document}